# Summation Problem Revisited - More Robust Computation

VACLAV SKALA
Department of Computer Science and Engineering
University of West Bohemia
Univerzitni 8
CZECH REPUBLIC
skala@kiv.zcu.cz   http://www.VaclavSkala.eu

*Abstract:* - Numerical data processing is a key task across different fields of computer technology use. However even simple summation of values is not precise due to the floating point representation use. This paper presents a practical algorithm for summation of values convenient for medium and large data sets. The proposed algorithm is simple, easy to implement. Its computational complexity is $O(N)$ in the contrary of the Exact Sign Summation Algorithm (ESSA) approach with $O(N^2)$ run-time complexity. The proposed algorithm is especially convenient for cases when exponent data differ significantly and many small values are summed with higher values.

*Key-Words:* - Numerical precision; floating point representation; linear algebra; matrix multiplication; summation; sign of sum; IEEE 758; robustness; stability; hashing.

## 1 Introduction

Numerical computation is used in many applications. Computer power is doubled every 18 months; memory capacity grows fast as well. However numerical representation is still restricted to floating point representation which has been standardized as IEEE 578 in 2008. There are many examples, how numerical non-robustness caused several disasters, e.g. North Sea Sleipner oil platform collapse, Tacoma Bridge collapse, Patriot missile failure, Ariana 5 rocket failure etc. All these catastrophes were somehow connected with numerical problems, instability and non-robustness of numerical computations.

However, there is no enough attention paid in education, engineering practice and software development. Robustness and precision of computation is becoming a challenging issue as with a growing computer power and memory capacity problems solved are becoming close to ill conditioned and huge data are to be processed. Nowadays, mostly 64 bit architecture is used enabling large memory and processing of large data sets, vectors or matrices. Tab.2 presents the IEEE 754-2008 standard. It should be noted that the standard specifies some special values, and their representation: positive infinity (+∞), negative infinity (−∞), a negative zero (−0) distinct from ordinary ("positive") zero, and "not a number" values (NaNs).

|  | Decimal usage | Binary usage |
|---|---|---|
| GigaByte [GB] | $10^9$ | $2^{30}$ |
| TeraByte [TB] | $10^{12}$ | $2^{30}$ |
| PentaByte [PB] | $10^{15}$ | $2^{30}$ |
| ExaByte [EB] | $10^{18}$ | $2^{30}$ |
| ZettaByte [ZB] | $10^{21}$ | $2^{30}$ |
| YottaByte [YB] | $10^{24}$ | $2^{30}$ |
|  |  |  |
| ???? | ?? | $2^{64}$ |

Table 1: Memory capacities

Unfortunately representations for rational numbers are limited to a single or double precision in many languages and quadruples or extended precisions are not generally supported by programming languages directly. It can be shown that for many even simple problems this is a severe limitation. Of course due to the precision of computation, there is a possibility to use an exact computation or interval arithmetic, but it leads to slow computations in general.

This paper presents a new approach to **Summation** and **Sign of Sum** problems that are fast, easy to implement.

| Name | Bits | Digits | | E min | E max |
|---|---|---|---|---|---|
| | | Exp | Mant. | | |
| Half | 16 | 5 | 10+1 | −14 | 15 |
| Single | 32 | 8 | 23+1 | −126 | 127 |
| Double | 64 | 11 | 52+1 | −1022 | 1023 |
| Extend. | 80 | 15 | 64+1 | −16382 | 16383 |
| Quad | 128 | 15 | 112+1 | −16382 | 16383 |

IEEE 754-2008 Standard
Table 2

## 2 Summation

Please, leave two blank lines between successive
There is a "very simple" problem, which is used in many textbooks, summation of a sequence of numbers, i.e.

$$S = \sum_{i=1}^{n} a_i \quad (1)$$

However, there are simple well known examples of summation incorrectness [15] (single precision is used):

$$S = \sum_{i=1}^{10^3} 10^{-3} = 0.999990701675415 \quad (2)$$

or

$$S = \sum_{i=1}^{10^4} 10^{-4} = 1.000053524971008 \quad (3)$$

It can be seen that in the both cases the result should be one, i.e. $S = 1$. The correctness in summation is very important in power series computations, matrix multiplications and solution of linear system of equations. The problem is even more complicated as results generally depend on the order of computation, e.g. how the values are actually order in the given data set.

$$\overline{S} = \sum_{i=1}^{10^6} \frac{1}{i} = 14.357357 \quad (4)$$

or if the reverse order is used

$$\underline{S} = \sum_{i=10^6}^{1} \frac{1}{i} = 14.392651 \quad (5)$$

If values $a_i$ are ordered we can get a slightly better estimation as $S = (\underline{S} + \overline{S})/2$. It means that even a very simple summation is not precise and reliable.

The problem gets even worse if summation is made for large interval of values. This example is not an "academic" problem one as it occurs in the matrix multiplication operation as today's matrices are quite large. The typical problem is the Fourier transform used in many computational packages and especially in physics and optics and it leads to large matrices, e.g. in digital holography [6], [9], radial basis function interpolation [14] or simplification of dynamic triangular meshes [16].

Another example of numerical imprecision is a computation of a function value. It is one of the basic common operations in engineering problems. However many programmers are not aware of the danger in the coding process. There seems to be two the most dangerous cases:

- division by a value close to zero, e.g. in an intersection computation of two nearly parallel lines
- addition or subtraction of two values with significantly different absolute value, e.g. recently mentioned $x^2 \pm y^2$ .

As the result of this, the summation (repeated addition) result depends on the order of summation in general.

Let us explore one very interesting case [9] and some other interesting comments [2], [10].

$$\begin{aligned} f(x,y) = 333.75y^6 \\ +x^2(11x^2y^2 - y^6 - 121y^4 - 2) \\ +5.5y^8 + x/(2y) \end{aligned} \quad (6)$$

The question is, what is the value of the function, if different floating point precisions are used and if it is evaluated at $x = 77617$, $y = 33096$.

$f = 6.33835 \; 10^{29}$  in single precision

$f = 1,1726039400532$  in double precision

$f = 1,17260394005317863185883490452 01838$
in extended precision

However even the result in the extended precision is incorrect and even the sign of the value itself is incorrect. The correct result is "somewhere" in the interval of

[−0,82739605994682136814116509547 98162920**05**,
−0,82739605994682136814116509547 9816 29**1986**]

if approx. 40 digits were used [10].

Of course this function is constructed in a special way, but it demonstrate that

- simple increase of precision does not guarantee the correctness of the result
- roundoff error has significant influence to for a limited floating point computation.

Detailed analysis of this function can be found in [2] and the correct result is

$$f(x,y) = -2 + \frac{x}{2y} = -\frac{54767}{66192} \quad (7)$$

Unfortunately precision of the numerical results is significantly influenced by compiler's properties and options used, as the optimization of the code is not considering the numerical stability issues but optimize the speed of computation.

### 2.1 Summation Problem
The summation is a computation of the sum of a sequence of values $a_1, a_2, \ldots, a_n$, i.e.

$$S = \sum_{i=1}^{n} a_i \quad (8)$$

From a mathematical point of view it is a trivial problem, from a programmer's naïve approach it is a simple sequence of code, however the problem gets complicated if floating point is used and values of $a_i$ differs in magnitude. The "Compensated summation" algorithm [3] [7] tries to solve this problem efficiently. Assuming $|a| \geq |b|$ then one computational step can be described as follows:

$$\tilde{s} := a + b\,;$$
$$\tilde{t} := \tilde{s} - b\,; \quad (9)$$
$$e = b - \tilde{t}\,;$$

It can be seen that $e$ is an error of computation as if $a$ and $b$ differs then part of the mantissa of $b$ is lost. From the "pure" mathematical point of view $e = 0$ as the sequence above is actually an identity, i.e. $e = b - ((a+b) - 1)$ and $a + b = \tilde{s} + e$.

However it should be noted that if

$$\sum_{i=1}^{n} |a_i| \gg \left|\sum_{i=1}^{n} a_i\right| \quad (10)$$

then the relative error of the summation cannot be guaranteed [7,p.3].

### 2.2 The Compensated Sum algorithm

**procedure** Compensated_Sum
(**Input**: **vector double** a[1:n], **int64** n;
**Output**: **double** S, error);
{ **double**: temp, q;
   S := 0; error := 0;
   **for** i := 1 **to** n **do**
   { temp := S;
     q := a[i] + error;
     # compensated sum #
     S := temp + q;
     # cumulative sum #
     error := (temp - S) + q;
     # error estimate #
   }
} # end of Compensated_Sum #

Algorithm 1

The values of the **a** vector **have to be ordered** which requires $O(n \lg n)$ complexity and several operations are used instead of one, i.e. instead of $s := s + a_i$ only. Unfortunately, this approach is not practical for large data sets, e.g. for $n \gg 10^5$.

## 3 Non-Traditional Floats Operations
Current programming languages offer standard data types and numerical operations including floating point representation. The IEEE 578 floating point representations, see Tab.2, are highly optimized, now. However some operations very often used in computational practice are not implemented efficiently in many programming languages. The influence of basic numerical operations to a numerical precision is well known and can be formalized using the interval arithmetic as follows.

### 3.1 Precision of Operations
Let as assume that we have two numbers $x$ and $y$, where: $x \in [a, b]$, $y \in [c, d]$, i.e. all values between $a$ and $b$ are represented by the same value in the floating point representation. The following interpretation of the basic arithmetic operations demonstrate how the actual precision is defined.

- $x + y \in [a + c, b + d]$
- $x - y \in [a - d, b - c]$
- $x \times y \in [min(ac, ad, bc, bd),$
  $\qquad\qquad max(ac, ad, bc, bd)]$
- $x / y \in [min(a/c, a/d, b/c, b/d), max(a/c, a/d, b/c, b/d)]$ if $y \neq 0$

The division operation is the longest one, except of comparison operation, and the worst one from the precision point of view.

However, the operations like "divide by 2" or "multiply by 2" often used are mostly translated to a binary code as a general division or multiplication operation. Therefore there is a natural question how it could be implemented efficiently and robustly. It should be noted that optimization made by a compiler is targeted to optimization of speed of computation regardless to precision and robustness. This might be very dangerous in some applications.

Let us introduce the **union** construction. It is actually well known **equivalence** construction from the Fortran programming language which enables to use the same referred memory differently. The construction can be described for our purposes as follows:

**union** (**double**: d; **uint16** aa, bb, cc, dd): q;
  # in the case of double precision #
resp.

**union** (**float**: d; **uint16** aa, bb): q;
  # in the case of single precision #

It actually means that memory element $q$ can be seen as a **double** or as 4 consecutive memory elements of **uint16**, resp. as 2 consecutive memory elements of **uint16** in the case of a **float**, i.e. in the **single** precision case. This construction, i.e. bit manipulation with exponent or mantissa, is considered as "dangerous" but it is actually very useful one.

### 3.2 Exponent Extraction
In some computational cases we need to extract the binary exponent EXP of the given value $q$ the **union** construction can be used efficiently.

General construction for a single or double precision is defined as

EXP := ((q.aa **land** MASK) **shr** m) **land** MASK_1;
  # **land** not needed if **uint** is unsigned **int** #
where: **land** is bitwise and operation, **shr** is shift right, MASK is a binary 16 bits mask and $m$ is the argument for the shift operation. In this way we can manipulate with the exponent, i.e. read or rewrite it. It should be noted that the exponent value of the given of a value in the floating point representation is stored as a binary exponent with a shifted zero and MASK_1 removes the mantissa bit.

The sequence above is simple, but special cases are not considered. Tab.3 presents corresponding values for different precision and

$$\text{Mask\_2} = \sim \text{MASK} \tag{11}$$

where ~ means bitwise negation.

### 3.3 Exponent Change
It is useful to have also operation for the exponent change, resp. it's rewriting. This is potentially dangerous operation as the exponent is to be overwritten. Therefore it is a little bit complicated operation as some attention must be paid to the correctness of the implementation. Note that the following construction for the exponent rewriting by a new exponent value EXP does not handle all the possible cases, but it is valid in principle.

q.aa := (q.aa **land** MASK_2) **lor** (EXP **shl** m)

where: **shl** operation means shift left.

Now, an efficient implementation of the floating point division by 2 and multiplication by 2 is becoming simple and fast.

### 3.4 Division and Multiplication by 2 in the Floating Point Representation
The division operation leads to principal loss of precision and to instability in general. Therefore there is a legitimate question if and why imprecise general division operation must be used. The same question is also valid for multiplication operation.

Analyzing the floating point representation it is clear that the exponent value EXP is **decreased** by 1 in the case of the division operation by 2, while in the case of multiplication by 2, the value EXP is **increased** by 1. Of course the "underflow" and "overflow" of the exponent has to be checked.

| Precision | MASK | MASK_1 | MASK_2 | m | | Exp_Val |
|---|---|---|---|---|---|---|
| Single | &7F80 | &00FF | &807F | 7 | | 255 |
| Double | &7FF0 | &07FF | &800F | 4 | | 2047 |

Table 3: Mask for non-traditional operations with floating numbers

The above presented operations can be used for a new summation algorithm based on bucketing of exponents and for simple but not exact **Sign of Sum** problem solution, which is fast and gives good precision, as well. As the interval of exponents, see tab.3, is relatively small in relation to an expected number of items to be processed (we expect large sets to be processed), additional memory requirements and processing time are acceptable. The given approach can be used easily for more precise large matrix multiplication, too.

## 4 Sign of Sum Problem

There is a well known Sign of Sum problem, which solves exactly the problem [11], [12]. However, the values have to be sorted, which is of $O(n \lg n)$ complexity and the algorithm itself is of $O(n^2)$ complexity [12], [5], which is not acceptable for large data sets. The algorithm relies on splitting input to two groups positive are stored in **A** while negative values are stored in **B** and final sum is then computed as

$$S = A - B = \sum_{i=1}^{k} a_i - \sum_{j=1}^{l} b_j \quad (12)$$

### 4.1 Exact Sign of Sum Algorithm (ESSA)

Input: $a_i$, $b_j$ (i = 1, ... , k, j = 1, . . . , ℓ - renamed inputs $a_{k+i}$ as $-b_i$ , i = 1, ... , ℓ = n−k)

Output: The sign of S = $S = \sum a_i - \sum b_i$

1. (BASIS) Terminate with the correct output in the following cases:
    1.1 **if** k = ℓ = 0 **then** S = 0.
    1.2 **if** k > ℓ = 0 **then** S > 0.
    1.3 **if** ℓ > k = 0 **then** S < 0.
    1.4 **if** $a_1 \geq ℓ\, 2^{F1+1}$ **then** S > 0.
    1.5 **if** $b_1 \geq k\, 2^{E1+1}$ **then** S < 0.

2. (AUXILIARY VARIABLES)
    a′ = a″ = b′ = b″ = 0;
3. (PROCESSING THE LEADING SUMMANDS)
    **CASE** E1 = F1:
        If $a_1 \geq b_1$ then a′ ← $a_1 - b_1$
        Else b′ ← $b_1 - a_1$;
    **CASE** E1 > F1:
        u ← $2^{F1+1}$;
        a′ ← $a_1 - u$, a″ ← $u - b_1$;
    **CASE** E1 < F1:
        v ← $2^{E1+1}$;
        b′ ← $b_1 - u$, b″ ← $u - a_1$;
4. (UPDATE BOTH LISTS)
    Discard $a_1$, $b_1$ from list.
    Insert a′, a″ into the a-list (only non-zero values need to be inserted).
    Insert b′, b″ into the b-list (only non-zero values need to be inserted).
    Update the values of k, ℓ and return to Step 1.

Algorithm 2
(Taken from [5, lecture 4, p6])

It can be seen that the algorithm is quite complex and requires data ordering as well. However, there are many cases when we do not need EXACT computation, but as precise computation as possible. In this case a modified summation algorithm based on hashing can be used with high computational efficiency. Sign of Sum algorithm precision is The summation problem for large data sets is becoming quite difficult. Let us consider the case, when the number of values $n \gg 10^6$ and we do not want to use ESSA (Exact Sum Sign Algorithm) approach due to its complexity. Sorting values is not acceptable due to time response etc. The algorithm must be fast and easy to implement as large data sets are to be processed. In the following double floating point representation will be used, unless otherwise noted.

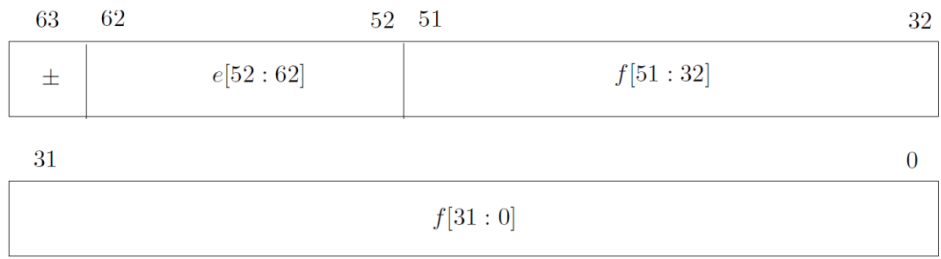

Figure 1: IEEE floating number representation

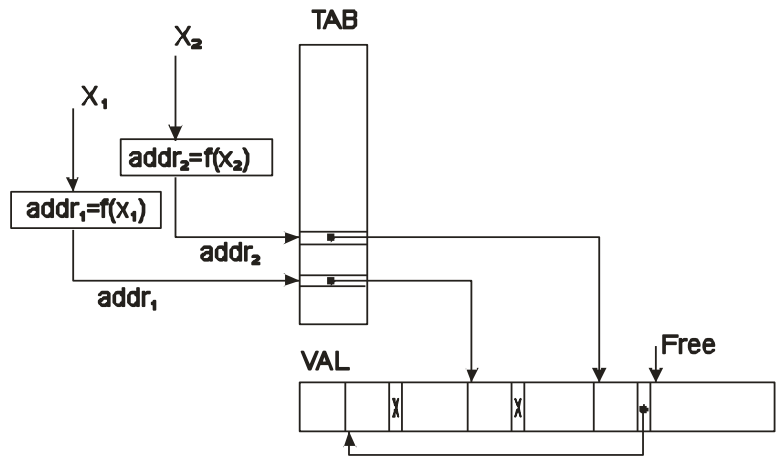

Figure 2: Hash function data structure in principle

In the double precision case the mantissa is represented by 52 bits, while exponent is represented by 11 bits with the shifted zero, i.e. if EXP = 0 then a value stored actually is $mantissa * 2^{-1024}$ (in reality the *mantissa* has 1 "invisible" bit more).

As there are only 2048 values for EXP, the EXP value can be considered as the address to the hash table, where the cumulative sum for each binary exponent value is stored. Now, the **Hash Summation** algorithm can be described as follows: important in geometry when long and thin shapes are processed, e.g. in surface extraction from implicitly defined objects [1] [16].

## 4 Hashing techniques

Hashing technique is used in many applications [8], [13], [15]. One of the advantages is that retrieval complexity is $O_{exp}(1)$ if a good hash function is used regardless number of items stored, i.e. design of the hash function is critical to run-time efficiency. Principle of the hashing technique is presented in Fig.2.

In the case of the Summation algorithm application of the hash function is quite simple. The value of the $i^{th}$ input element, i.e. $a_i$, is stored directly in the TAB table directly, while the address to the hash table is given by the binary exponent of the value stored. As the range of exponents is $\{-1022, \dots, 1023\}$ the hash table has 2048 positions only in the case of double precision, that acceptable length even for medium and large data sets.

## 5 Hash Based Sign Of Sum

The summation problem for large data sets is becoming quite difficult. Let us consider the case, when the number of values $n \gg 10^6$ and we do not want to use ESSA (Exact Sum Sign Algorithm) approach due to its complexity. Sorting values is not acceptable due to time response etc. The algorithm must be fast and easy to implement as large data sets are to be processed. In the following double floating point representation will be used, unless otherwise noted. In the double precision case the mantissa is represented by 52 bits, while exponent is represented by 11 bits with the shifted zero, i.e. if EXP = 0 then a value stored actually is $mantissa * 2^{-1024}$ (in reality the *mantissa* has 1 "invisible" bit more). As there are only 2048 values for EXP, the EXP value can be considered as the address to the hash table, where the cumulative sum for each binary exponent value is stored. Now, the **Hash Summation** algorithm can be described as follows:

**procedure** Hash_Summation
(**In**: **double** a[1:n], **uint64** n; **Out**: **double** S);
{
# version for double precision #
# ---------------------------------------------------------- #
# Hardware constants for double IEEE 578/2008 #
# sign bit is to be eliminated #
# and part of the mantissa #
# MASK values &7FF0 for double #
# &7F80 for single precision #
# ---------------------------------------------------------- #
**const uint16** MASK = &7FF0;
**const uint16** Exp_Size = 2047, Exp_Zero = 1028;
**const int16** m=4;
**const int16** N=1000;  #max. level of recursion#

```
# ---------------------------------------------------------- #
uint16 addr, rek_lev;
double q_new;
double TAB[0 : Exp_Size];
uint LEVEL [0 : N]; #recursion levels #
type eqv =
    union (double: d; uint16: aa, bb, cc, dd);
    # a new type definition #

uint16 function Exponent (eqv val) inline;
{  return ( (val.aa and MASK) shr m)
# mask and shift to the right #
};
# ----------- RECURSIVE CASE -------- #
procedure ADD (eqv val; uint16 exp_val);
{   eqv val_old,q;
    uint16 exp_q;
    val_old := TAB[exp_val];
    TAB[exp_val] := 0;
    q := val + val_old;  # new value #
    exp_q := Exponent(q); # new value exponent #
    # if exponents are equal then store #
    # the value directly else call ADD recursively #
    if exp_q = exp_val then
    {   TAB[exp_q] := val;
        LEVEL[rek_lev] +:= 1;
        rek_lev:=0
    }
    else
    {   rek_lev +:= 1;
        ADD(q, exp_q)
    }
};

    # TAB initialization #
    for i := 0 to Exp_Size do
        TAB[i] := 0;
    for i := 0 to N do
        LEVEL[i] := 0;

    for i : = 1 to n do
    {   q := a[i]; # conversion to the eqv type #
        rek_lev := 0;
        ADD (q, Exponent(q));
    };
    # Final summation #
    S := TAB[Exp_Zero];
    # the lowest exponents first #
    for i : = 1 to Exp_No do
    {   S+:=TAB[Exp_Zero+i]+TAB[Exp_Zero-i];
    };
} # end of Hash_Summation #
```

Recursive algorithm

Algorithm 2

However, this algorithm is recursive that leads to a higher computational complexity due to subsequent recursive calls. Simple non-recursive modification stores values directly in the TAB table which leads to cases where values are stored in non-exact positions in the TAB table; position can differ by $\pm k$, where k=1 or k=2 usually.

```
# ----------- NON-RECURSIVE CASE -------- #
eqv: q;
double function GET (eqv q);
{   q.d := a[i];
    # position in TAB #
    addr := (q.aa and MASK) shr m;
    # mask and shift to the right #
    q_new := q.d + TAB[addr];
    # a new value of the partial sum #
};

for i : = 1 to n do
{   q.d := a[i];
    # position in TAB #
    addr := (q.aa and MASK) shr m;
    # mask and shift to the right #
    q_old := TAB[addr]; TAB[addr] := 0;
    q_new := q.d + TAB[addr];
    # a new value of the partial sum #
    TAB[addr] := q_new;
    # additional sequence should be placed here #
    # if code to be optimized #
};

    S := TAB[Exp_Zero];
    # the lowest exponents first #
    for i : = 1 to Exp_No do
    {   S+:=TAB[Exp_Zero+i]+TAB[Exp_Zero-i];
    };
# end of Hash_Summation #
```

Algorithm 3

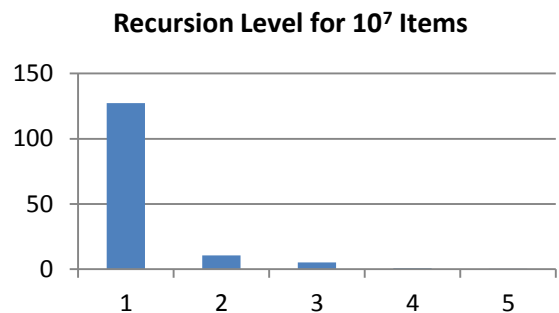

Typical example of the recursion level for $10^7$ items; No.of calls *1000

Figure 3

This is actually bucketing algorithm and can be easily improved as after some iterations the exponent of the value stored in the table TAB has higher value than expected and therefore should be stored in a different position in the table TAB. The additional sequence, showing the principle only, is simple but not optimal one:

```
# q_new–value not stored in the proper position? #
q.d := TAB[addr];
addr_new := (q.aa land MASK) shr m;
if addr_new ≠ adr then
# value actually stored has different exponent #
{   if addr_new >= 0
    or addr_new <= Exp_Val then
    # except of the 0 and Exp_Val exponent #
    {   TAB[addr] := 0.0;
        TAB[addr_new] := q.d;
    } ;
}
```

Algorithm 4

There should be an iterative "propagation" of the $q$ value up or down within the TAB to the correct position, but there is a small probability that the $q$ value would propagate repeatedly up or down. However, it will be partially corrected in the following steps of this algorithm. Also the code would be much more complicated and run-time efficiency would be lower.

## 6 Experimental Results

The proposed algorithms have been successfully tested with different functions, e.g.

$$\sum_{i=1}^{n} i = \frac{n(n+1)}{2} \tag{13}$$

$$\sum_{i=1}^{n} i^3 = \frac{n^2(n+1)^2}{4} \tag{14}$$

$$\sum_{i=1}^{\infty} \frac{1}{(i-1)!} = e \tag{15}$$

for $n > 10^7$ etc. For simplicity the single precision was used in experiments.

Also the following functions were tested

$$\sum_{i=1}^{n}(\ln(i+2) - \ln(i+1)) \tag{16}$$
$$= \ln(n+2) - \ln(2)$$

$$\sum_{i=1}^{n} \frac{1}{i(i+1)} = \sum_{i=1}^{n}\left(\frac{1}{i} - \frac{1}{i+1}\right) \tag{17}$$
$$= 1 - \frac{1}{n+1}$$

$$\sum_{i=1}^{\infty} \frac{1}{(i-1)!} = e \tag{18}$$

$$\sum_{i=1}^{n} 3\left(1 + \frac{i}{n}\right)^2 \frac{1}{n} = \frac{3}{n}\sum_{i=1}^{n}\left\{1 + \frac{2i}{n} + \frac{i^2}{n^2}\right\} \tag{19}$$
$$= \left\{6 + \frac{3}{n} + 1 + \frac{3}{2n} + \frac{3}{6n^2}\right\}$$

$$F(x_1, \ldots, x_n) = \sum_{i=1}^{n/2}(100(x_{2i} - x_{2i-1}^2)^2 + (1 - x_{2i})^2) \tag{20}$$

Extendex Rosenbrock function [3]

$$\lim_{n \to \infty} \sum_{i=1}^{n} \frac{1}{(2i-1)(2i+1)} = \frac{1}{2} \tag{21}$$

Table 7

## 7 Conclusion

A new approach to the Summation problem solution for large data sets has been presented. It is based on bucketing principle which helps to "re-sort" given values to buckets with the same binary exponent. The algorithm is not convenient for the cases when data have the similar binary exponent in the floating point representation, e.g. if data are from an interval $< 2^k, 2^{k+1} - 1 >$, as all data would fall into the same bucket.

## Acknowledgments
The author would like to express his thanks to students and colleagues at the University of West Bohemia in Plzen and VSB Technical University in Ostrava for their recommendations, constructive discussions and hints that helped to finish this work.

Many thanks belong to the anonymous reviewers for their valuable comments and suggestions that improved this paper significantly. This research was supported by the Ministry of Education of the Czech Republic – projects No.LH12181 and LG13047.